\documentclass{article}
\usepackage{graphics}
\usepackage{latexsym}

\def\6{\ensuremath{\langle}}
\def\9{\ensuremath{\rangle}}
\def\v{\ensuremath{{\mathbf{v}}}}
\def\p{\ensuremath{{\mathbf{p}}}}
\def\P{\ensuremath{{\mathbf{P}}}}
\def\k{\ensuremath{{\mathbf{k}}}}
\def\K{\ensuremath{{\mathbf{K}}}}
\def\x{\ensuremath{{\mathbf{x}}}}
\def\X{\ensuremath{{\mathbf{X}}}}
\newcommand{\mychoose}[2]{\left(\begin{array}{c}#1\\#2\end{array}\right)}

\begin{document}

\title{Information and Distinguishability of Ensembles of Identical Quantum States}

\author{Lev B. Levitin ({\tt levitin@bu.edu})\\ Tom Toffoli\protect\footnote{Partially supported by the Department of Energy under grant DE-FG02-99ER25414} \ ({\tt tt@bu.edu})\\ Zac Walton\protect\footnote{Supported by the National Science Foundation and the Boston University Photonics Center} \ ({\tt walton@bu.edu})\vspace{.15in}\\
Department of Electrical and Computer Engineering, Boston University\\
8 Saint Mary's Street, Boston, Massachusetts 02215}

%\\
%Department of Electrical and Computer Engineering, Boston University\\
%8 Saint Mary's Street, Boston, Massachusetts 02215

\maketitle 

\abstract{We consider a fixed quantum measurement performed over
$n$ identical copies of quantum states. Using a rigorous notion of
distinguishability based on Shannon's 12th theorem, we show that in the case
of a single qubit the number of distinguishable states is
$W(\alpha_1,\alpha_2,n)=|\alpha_1-\alpha_2|\sqrt{\frac{2n}{\pi e}}$, where
$(\alpha_1,\alpha_2)$ is the angle interval from which the states are chosen.
In the general case of an $N$-dimensional Hilbert space and an area $\Omega$ of
the domain on the unit sphere from which the states are chosen, the number of
distinguishable states is $W(N,n,\Omega)=\Omega(\frac{2n}{\pi
e})^{\frac{N-1}{2}}$.  The optimal distribution is uniform over the domain in
Cartesian coordinates.}

\section{Introduction}

In his remarkable 1981 paper, ``Statistical Distance and Hilbert
Space'' \cite{wootters}, Wootters showed that the statistical distance
between two vectors in Hilbert space is proportional to the angle
between these two vectors and does not depend on the position of the
vectors.  He defines statistical distance as the number of
distinguishable intermediate states between the two vectors.  However,
his notion of distinguishibility relies on the apparently arbitrary
criterion that two states are distinguishable if measurements
performed on $n$ identical copies of each state yield two
distributions whose means are separated by a constant factor times the
sum of the standard deviations of these distributions.  We use a more
rigorous notion of distinguishability based on Shannon's 12th
theorem \cite{shannon} and arrive at an expression for the number of
distinguishable states that is consistent with Wootters' result;
however, unlike that result, our expression does not depend on an
arbitrary choice of the distinguishability criterion.  Rather, our
notion of distinguishibility is predicated on the guarantee that the
measurer be able to distinguish between the quantum states with
probability approaching $1$ as the number $n$ of copies of identical
states in a sample tends to infinity.  Wootters shows that for large
$n$ the number of distinguishable states between the vectors
$\alpha_1$ and $\alpha_2$ is proportional to $|\alpha_1-\alpha_2|\sqrt
n$, where $\alpha$ is the angle of the vector from some reference
direction in the plane spanned by the two vectors.  We show in Section 2 that
the actual number of distinguishable states in a 2-dimensional Hilbert space is
\begin{equation}
\mbox{W}(\alpha_1,\alpha_2,n)=e^{\mbox{{\footnotesize I}}_{\mbox{\scriptsize sup}}(P;K)}
=|\alpha_1-\alpha_2|\sqrt{\frac{2n}{\pi e}}
\end{equation}
 where $\mbox{I}_{\mbox{\scriptsize sup}}(P;K)$ is the maximum mutual information between the (random)
quantum state and the results of measurements.  We prove that this
maximum is achieved for an ensemble of quantum states with the uniform
distribution of the angle $\alpha$ for any interval 
$[\alpha_2,\alpha_1]$.  The
independence of the number of distinguishable states of the position
of the interval $[\alpha_2,\alpha_1]$ is a remarkable asymptotic 
property
that does not hold for small values of $n$ (cf. \cite{levitin}).

Section 3 of this paper provides a generalization of these results to the case of an $N$-dimensional Hilbert space of states of the quantum system.  It turns out that the number of distinguishable states depends only on the area $\Omega$ of the domain on the unit sphere from which the states can be chosen, but does not depend on the shape and position of this domain.  The optimal distribution is uniform over this domain in Cartesian coordinates, and the number of distinguishable states is $W(N,n,\Omega )=\Omega (\frac{2n}{\pi e})^{(N-1)/2}$.
\section{The Case of a Single Qubit}
\subsection{Formulation of the Problem}

\begin{figure}[ht]
\begin{center}
\includegraphics{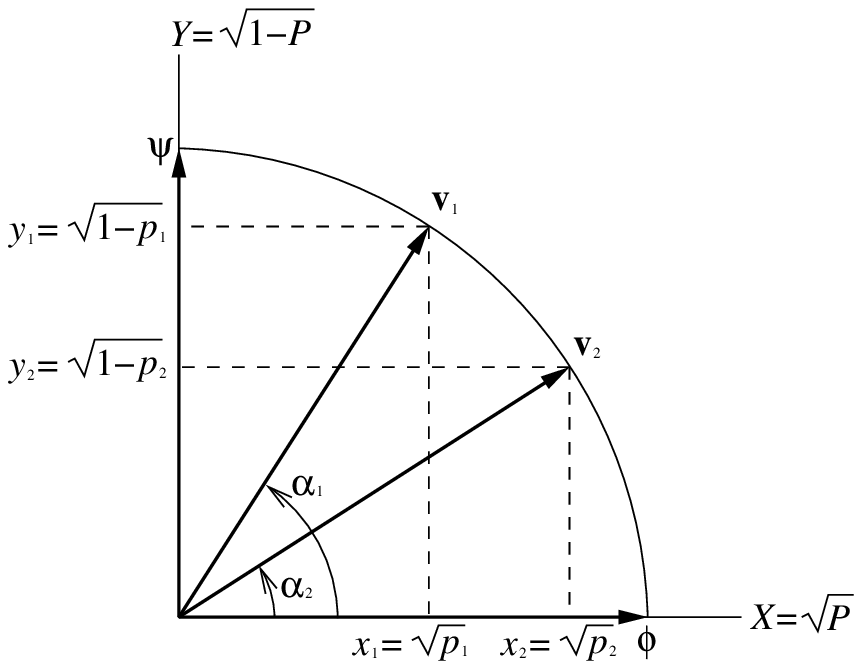}
\caption{The two state vectors $\v_1$, $\v_2$ and their projections on the the basis elements ${\scriptstyle {\mathsf{\Phi}}}$ and ${\scriptstyle {\mathsf{\Psi}}}$.}
\label{fig}
\end{center}
\end{figure}

Consider a quantum physical system whose states are unit vectors in a
$2$-dimensional complex Hilbert space $\mbox{\bf{C}}^2$ (the so-called
``qubit''). Denote the state vector by \v\ and let
$({\scriptstyle {\mathsf{\Phi}}},{\scriptstyle {\mathsf{\Psi}}})$ be an orthonormal basis in the Hilbert space, so that $\v=a{\scriptstyle {\mathsf{\Phi}}} +
b{\scriptstyle {\mathsf{\Psi}}}$, where $a=\6\v|{\scriptstyle {\mathsf{\Phi}}}\9$,
$b=\6\v|{\scriptstyle {\mathsf{\Psi}}}\9$ are inner products and $|a|^2+|b|^2=1$.  Then
$|a|^2=p$ and $|b|^2=1-p$ are probabilities of two possible outcomes
of the measurement performed over the state \v\ in the $({\scriptstyle {\mathsf{\Phi}}},{\scriptstyle {\mathsf{\Psi}}})$
basis.  Obviously, these probabilities do not depend on the phases of
the coefficients $a$ and $b$, and, therefore, all quantum states with
the same magnitudes $|a|=x$ and $|b|=y$ are indistinguishable by this
measurement.  Hence, the state space $\mbox{\bf{S}}^3$ can be reduced to the
non-negative quadrant of a circle in a real $2$-dimensional Euclidean
space (Fig.~\ref{fig}), spanned by ${\scriptstyle {\mathsf{\Phi}}}$ and ${\scriptstyle {\mathsf{\Psi}}}$.  Now let $\v_1$
and $\v_2$ be two distinct state vectors, such that
\begin{equation}
\v_i=x_i{\scriptstyle {\mathsf{\Phi}}}+y_i{\scriptstyle {\mathsf{\Psi}}}\ \ \mbox{where}\ \ x_i=\sqrt{p_i}\ \ \mbox{and}\ \ y_i=\sqrt{1-p_i},\ \ \mbox{for}\ \ i=1,2\,.
\end{equation}
Denote by $\alpha_i$ the angle between ${\scriptstyle {\mathsf{\Phi}}}$ and $\v_i$, so that
\begin{equation}
p_i=\cos^2{\alpha_i},\ \ 1-p_i=\sin^2{\alpha_i},\ \ i=1,2\,.
\end{equation}
 
Suppose, we want to distinguish between various quantum states chosen
from the interval of angles $[\alpha_2,\alpha_1]$ by performing
measurements in the $({\scriptstyle {\mathsf{\Phi}}},{\scriptstyle {\mathsf{\Psi}}})$ basis.  Further, assume that we are allowed to perform the measurement over $n$ identical copies of each quantum state.

{\bf Problem:} What determines the number of distinguishable states,
and what is the asymptotic expression for the number of states in the
interval $[\alpha_2,\alpha_1]$ that can be distinguished with
probability approaching $1$ when $n$ tends to infinity?

As shown in the next section, the problem can be rigorously analyzed by applying concepts and results of Shannon's information theory.

\subsection{Information-Theoretical Description}

Suppose the state vectors are chosen from the angle interval
$[\alpha_2,\alpha_1]$ with certain probability density function (p.d.f.)
$\mbox{P}_{\scriptstyle {\textsf A}}(\alpha)$, where ${\textsf A}$ is a random variable that takes on values
from $[\alpha_2,\alpha_1]$, $\alpha\in[\alpha_2,\alpha_1]$. Let $\mbox{P}_P(p)$
be the p.d.f. of the random variable $P$ that takes on values $p$, where $p$ is the probability of the state vector to be projected as the result of the measurement onto basis vector ${\scriptstyle {\mathsf{\Phi}}}$.  Obviously, $P=\cos^2{\textsf A}$, and the value of $P$ (or of ${\textsf A}$) characterizes uniquely the chosen quantum state.  In a series of $n$ measurements, let $K$ be the (random) number of measurements which have resulted in projectios onto ${\scriptstyle {\mathsf{\Phi}}}$. The conditional probability distribution of $K$ given $P$ is binomial:
\begin{equation}
\mbox{P}_{K/P}(k/p)=\mychoose{n}{k}p^k(1-p)^{n-k},\ \ \mbox{where}\ \ k=0,1,\dots,n\,.\label{cond-prob}
\end{equation}

The values of $K$ obtained in the measurement are the only data available from which one can infer about the value of $P$, i.e., about the choice of a quantum state.

Let $\mbox{P}_K(k)$ be the marginal probability distribution of $K$.  The
information $\mbox{I}(K;P)$ in $K$ about $P$ is given by 
\begin{equation}
\mbox{I}(K;P)=\int^{p_2}_{p_1}\!\sum^n_{k=0}\mbox{P}_P(p)\mbox{P}_{K/P}(k/p)\ln\frac{\mbox{P}_{K/P}(k/p)}{\mbox{P}_K(k)}\,dp\,.\label{information}
\end{equation}

The importance of considering information $\mbox{I}(K,P)$ stems from Shannon's 12th theorem \cite{shannon} which, for our setting of the problem, can be rephrased in the following way. 

Let $\mbox{S}=\{\p\}$, where $\p$ is an $n$-dimensional vector
$\p=(p,p,\ldots,p)$ and $p\in[p_1,p_2]$ be the set of all possible input
signals and $\mbox{\bf{Z}}_n=\{0,1,\ldots,n\}$ be the set of all output
signals in a communication channel with a conditional probability
distribution given by (\ref{cond-prob}).  Let $L$ be the length of a sequence
of such input signals used independently. Then for any $\varepsilon\!>\!0$
the maximum number $\mbox{M}(L,\varepsilon)$ of input signals that can be
chosen from $\mbox{S}$ in such a way that the probability of error (incorrect
decision about \p\ based on the value of the output signal
$k\in\mathcal{Z}_n$) does not exceed $\varepsilon$ satisfies the asymptotic
property:
\begin{equation}
\lim_{L\rightarrow\infty}\left[\frac{\ln{\mbox{M}(L,\varepsilon)}}{L}\right]={\mbox{I}_{\mbox{\scriptsize sup}}(K;P)},
\end{equation}
where $\mbox{I}_{\mbox{\scriptsize sup}}(K;P)$ is the least upper bound of $\mbox{I}(K;P)$ given by (\ref{information}) over all possible probability distributions $\mbox{P}_P(p)$ of the input parameter $P$.

Note that the asymptotic expression for $\mbox{M}(L,\varepsilon)$ in
fact does not depend on $\varepsilon$.  This means that the number of distinct input signals (different
values of $P$) that can be distinguished with probability arbitrarily
close to $1$ is $e^{\mbox{{\footnotesize I}}_{\mbox{\scriptsize sup}}(K;P)}$.  The problem
is reduced now to the computation of $\mbox{I}_{\mbox{\scriptsize sup}}(K;P)$ under the
condition that $P$ takes on values in $[p_1,p_2]$.  This problem is
very difficult, in general.  However, the following important theorem
will be helpful.

Define individual information in $P=p$ about $K$ as 
\begin{equation}
\mbox{I}(K;p)=\sum_{k=0}^n\mbox{P}_{K/P}(k/p)\ln\frac{\mbox{P}_{K/P}(k/p)}{\mbox{P}_K(k)}\,.
\end{equation}
As is well known (e.g. \cite{mansuripur}), $\mbox{I}(K;P)$ achieves the maximum value $\mbox{I}_{\mbox{\scriptsize sup}}(K;P)$ for such a distribution $\mbox{P}_P(p)$ that there exists a constant $\mbox{I}$ such that 
\begin{equation}
\mbox{I}(K;p)=\mbox{I}\ \ \mbox{for all}\ \ p\ \ \mbox{such that}\ \ \mbox{P}_P(p)>0
\end{equation}
and
\begin{equation}
\mbox{I}(K;p)<\mbox{I}\ \ \mbox{for all}\ \ p\ \ \mbox{such that}\ \ \mbox{P}_P(p)=0\,.
\end{equation}
Then $\mbox{I}_{\mbox{\scriptsize sup}}(K;P)=\mbox{I}$.

\subsection{The Number of Distinguishable States}

When $n$ is large, the binomial distribution (\ref{cond-prob}) can be well-approximated by a Gaussian distribution:
\begin{equation}
\mbox{P}_{K/P}(k/p)=\mychoose{n}{k}p^k(1-p)^{n-k}\approx\frac{1}{\sqrt{2\pi p(1-p)n}}e^{-\frac{(k-pn)^2}{2p(1-p)n}}\,. \label{gauss-dist}
\end{equation}

For large $n$, distribution (\ref{gauss-dist}) has a very sharp maximum at $k=pn$, so that the Laplace method \cite{debruijn} can be used for evaluation of integrals involving (\ref{gauss-dist}).

Consider a uniform distribution over the angle interval $[\alpha_2,\alpha_1]$,
\begin{equation}
\mbox{P}_{\scriptstyle {\textsf A}}(\alpha)=\frac{1}{|\alpha_1-\alpha_2|}\ \ \mbox{for}\ \ \alpha\in[\alpha_2,\alpha_1]\,. \label{alpha-prob}
\end{equation}

The corresponding distribution of the probability $P$ is
\begin{equation}
\mbox{P}_P(p)=\mbox{P}_{\scriptstyle {\textsf A}}(\alpha)\left|\frac{d\alpha}{dp}\right|=\frac{\left[p(1-p)\right]^{-1/2}}{2|\alpha_1-\alpha_2|}\ \ \mbox{where}\ \ \alpha_i=\cos^{-1}{\sqrt{p_i}}\,,\,i=1,2\,.\label{p-prob}
\end{equation}

We will prove that for large $n$ this distribution yields the maximum of $\mbox{I}(K;P)$. 
The marginal probability distribution $\mbox{P}_K(k)$ can be evaluated as follows (assuming $p_2>p_1$):
\begin{eqnarray}
\mbox{P}_K(k) &= & \int^{p_2}_{p_1}\mbox{P}_P(p)\mbox{P}_{K/P}(k/p)\,dp \nonumber \\
& \approx &\frac{1}{2|\alpha_1-\alpha_2|}\int_{p_1}^{p_2}\frac{1}{p(1-p)\sqrt{2\pi n}}e^{-\frac{(k-np)^2}{2p(1-p)n}}\,dp\,.
\end{eqnarray}
If the point of maximum $p=\frac{k}{n}$ of the exponential function in
the integrand is within the interval $[p_1,p_2]$, the integration interval can be extended to $(-\infty,\infty)$.  Otherwise, the value of the integral approaches zero when $n$ tends to infinity.  Thus, for large $n$ we obtain: 
\begin{equation}
\mbox{P}_K(k)\approx\left\{\begin{array}{ll}\frac{1}{2|\alpha_1-\alpha_2|\sqrt{k(n-k)}} & \mbox{if}\ \ np_1\le k\le np_2\\
0 & \mbox{otherwise.}\end{array}\right.
\end{equation}
Note that, as could be expected, the distribution of $K$ for large $n$ is the discrete counterpart of the distribution of $P$.  Now we can evaluate the individual information $\mbox{I}(K;p)$.
\begin{eqnarray}
\mbox{I}(K;p) & \approx & \sum^{\lfloor np_1\rfloor}_{k=\lceil np_1 \rceil} \mbox{P}_{K/P}(k/p)\ln\frac{\mbox{P}_{K/P}(k/p)}{\mbox{P}_K(k)} \nonumber \\
& \approx & \int_{np_1}^{np_2}\frac{dk}{p(1-p)\sqrt{2\pi n}}e^{-\frac{(k-np)^2}{2p(1-p)n}}\left[\ln\mbox{P}_{K/P}(k/p)-\ln\mbox{P}_K(k)\right] \nonumber \\
& \label{approx}
\end{eqnarray}

The first term in (\ref{approx}) is the differential entropy of a Gaussian distribution (with the opposite sign), the second one can be evaluated by the Laplace method.  Hence, asymptotically,
\begin{eqnarray}
\mbox{I}(K;p) & = & -\frac{1}{2}\ln[2\pi e p(1-p)n] + \frac{1}{2}\ln[p(1-p)n^2] + \ln 2|\alpha_1-\alpha_2| \nonumber \\
& = & \frac{1}{2}\ln\frac{2n}{\pi e} +\ln|\alpha_1-\alpha_2|
\end{eqnarray}

Note that $\mbox{I}(K;p)$ takes on the same value for any $p\in[p_1,p_2]$.  Hence, distribution (\ref{alpha-prob}) (or (\ref{p-prob})) is the optimal one for large $n$, and the maximum information $\mbox{I}_{\mbox{\scriptsize sup}}(K;P)$ is expressed asymptotically as given below.
\begin{equation}
\mbox{I}_{\mbox{\scriptsize sup}}(K;P)=\frac{1}{2}\ln\frac{2n}{\pi e}+\ln|\alpha_1-\alpha_2|
\end{equation}
Thus, the number of distinguishable quantum states in the interval of angles $[\alpha_2,\alpha_1]$ is proportional to the length of the interval and to $\sqrt{n}$.  It does not depend on the position of the interval in the circle.
\begin{equation}
\mbox{W}(n,\alpha_1,\alpha_2)=e^{\mbox{{\footnotesize I}}_{\mbox{\scriptsize sup}}(K;P)}=|\alpha_1-\alpha_2|\sqrt\frac{2n}{\pi e} \label{final-expression}
\end{equation}

Of course, the range of ${\textsf A}$ may consist of several separated intervals.
Then (\ref{final-expression}) remains valid, as long as $n$ is
sufficiently large, so that each interval has many distinguishable
states; also, $|\alpha_1-\alpha_2|$ should be replaced by the total
length of the intervals.

For given $n$, (\ref{final-expression}) achieves maximum if $|\alpha_1-\alpha_2|=\pi/2$. Hence, 
\begin{equation}
\mbox{W}_{\mbox{\scriptsize max}}(n)=\sqrt{\frac{\pi n}{2 e}}.
\end{equation}

\section{The $N$-Dimensional Case}
Consider now a quantum system whose states are unit vectors in an
$N$-dimensional complex Hilbert space $\mbox{\bf{C}}^N$.  Choose an
orthogonal basis in $\mbox{\bf{C}}^N$ corresponding to a direct (von Neumann)
measurement.  Since all quantum states having the same projections on the
basis vectors are indistinguishable by this measurement, the state space
$\mbox{\bf{S}}^{2N-1}$ is reduced to the non-negative orthant of the unit
sphere $\mbox{\bf{S}}^{N-1}$ in the real $N$-dimensional Euclidean space
$\mbox{\bf{R}}^N$.  Each state vector is described now by $N$ Cartesian
coordinates $\x=(x_1,x_2,\ldots ,x_N)$, $\sum_{i=1}^N{x_i}^2=1$, and
$p_i={x_i}^2$ is the probability of the $i$-th outcome of the
measurement. Suppose we want to distinguish between states chosen from a
domain $\mbox{\bf{D}}$ of the non-negative orthant of $\mbox{\bf{S}}^{N-1}$,
and assume we are allowed to perform the same measurement over $n$ identical
copies of each quantum state, where $n\gg 1$.  Let the quantum states be
chosen with probability density function (p.d.f.) $\mbox{P}_{\P}(\p)=\mbox{P}_{\P}(p_1,\ldots ,p_N)$, where $\sum_{i=1}^Np_i=1$.  The outcome of such a measurement performed over $n$ identical states is an $N$-dimensional random variable $\K$ which takes on values $\k=(k_1,k_2,\ldots ,k_N)$, where $k_i$ ($i=1,2,\ldots , N)$ is the number of cases when the $i$-th result has been obtained.  The conditional probability distribution of $\K$ given $\P$ is multinomial:
\begin{equation}
\mbox{P}_{\K/\P}(k_1,\ldots ,k_m/p_1,\ldots ,p_m)=\frac{n!}{\prod_{i=1}^N{k_i}!}\prod_{i=1}^N{p_i}^{k_i},\label{multinomial}
\end{equation}
where $\sum_{i=1}^Nk_i=n$. 

Denote by $\mbox{P}_{\K}(\k)$ the marginal probability distribution of $\K$.  Then the information $\mbox{I}(\K;\P)$ in $\K$ about $\P$ is given by an expression similar to (\ref{information}):
\begin{equation}
\mbox{I}(\K;\P)=\int_{\p\in\mbox{\bf{D}}}\!\sum_{\k}\mbox{P}_\P(\p)\mbox{P}_{\K/\P}(\k/\p)\ln\frac{\mbox{P}_{\K/\P}(\k/\p)}{\mbox{P}_{\K}(\k)}\,d\p\,,\label{information-general}
\end{equation}
where summation is taken over all $\k$ such that $\sum_{i=1}^Nk_i=n$.

It follows from Shannon's 12th theorem that for any $\varepsilon>0$ the maximum number of distinct states $\mbox{W}(N,n,\varepsilon)$ chosen from $\mbox{\bf{D}}$ in such a way that the probability of incorrect identification of the state based on the results $\K$ of the measurement does not exceed $\varepsilon$ satisfies the limit
\begin{equation}
\lim_{N\rightarrow\infty}\lim_{n\rightarrow\infty}\frac{\ln{\mbox{W}(N,n,\varepsilon)}}{{\mbox{I}_{\mbox{\scriptsize sup}}(K;P)}}=1.
\end{equation}
Here $\mbox{I}_{\mbox{\scriptsize sup}}(\K;\P)$ is the least upper bound of $\mbox{I}(\K;\P)$ over all possible $\mbox{P}_\P(\p)$. Note that, in contrast with the 2-dimensional case, there is no need to consider sequences of distinct states provided $n$ and $N$ are sufficiently large.

Thus the number of distinct states (different values of $\P$) that can be distinguished with probability arbitrarily close to 1 is given by $e^{\mbox{I}_{\mbox{\scriptsize sup}}(\K;\P)}$.  The computation of $\mbox{I}_{\mbox{\scriptsize sup}}(\K;\P)$ can be performed along the same lines as in the 2-dimensional case.  For large $n$ ($n/N\gg 1$), the multinomial distribution (\ref{multinomial}) can be approximated by the $N$-dimensional Gaussian distribution \cite{gnedenko}
\begin{equation}
\mbox{P}_{\K/\P}(\k/\p)\approx\frac{e^{-\frac{1}{2}\sum_{i=1}^N\frac{(k_i-p_in)^2}{p_in}}\delta(\sum_{i=1}^Nk_i-n)}{(2\pi n)^{\frac{N-1}{2}}\prod_{i=1}^N{p_i}^{\frac{1}{2}}}\label{gnedenko-eqn}
\end{equation}
Consider the distribution $\mbox{P}_{\X}(\x)$ of the states which is uniform over the domain $\mbox{\bf{D}}$.  Denote the area of $\mbox{\bf{D}}$ by $|\mbox{\bf{D}}|=\Omega$. Then
\begin{equation}
\mbox{P}_{\X}(\x)=\frac{1}{\Omega}\,{\delta\left(\sqrt{\sum_{i=1}^N{x_i}^2}-1\right)}\label{delta}
\end{equation}
for $\x\in\mbox{\bf{D}}$, and $\mbox{p}_{\X}(\x)=0$ otherwise.  We will show that for large $n$ this distribution yields the maximum of $\mbox{I}(\K;\P)$.
Distribution (\ref{delta}) corresponds to the following distribution of the random variable $\P$ over the domain $\mbox{\bf{D}}$:
\begin{eqnarray}
\mbox{P}_{\P}(p_1,\ldots , p_N)& =&\frac{1}{\Omega}\mbox{J}\left(\frac{x_1,\ldots,x_N}{p_1,\ldots,p_N}\right)\delta\left(\sqrt{\sum_{i=1}^Np_i}-1\right)\\
& = & \frac{\prod_{i=1}^N{p_i}^{-\frac{1}{2}}\delta\left(\sum_{i=1}^Np_i-1\right)}{2^{N-1}\Omega}
\end{eqnarray}
where $\mbox{J}\left(\frac{x_1,\ldots,x_N}{p_1,\ldots,p_N}\right)$ is the Jacobian of the coordinate transformation from $\x$ to $\p$.
The marginal probability distribution of $\K$ is given by
\begin{equation}
\mbox{P}_{\K}(k_1,\ldots,k_N)=\int_{\mbox{\bf{D}}}\mbox{P}_{\P}(\p)\mbox{P}_{\K/\P}(\k/\p)dp_1\ldots dp_N.\label{integrand}
\end{equation}
For large $n$, the integrand in (\ref{integrand}) has a sharp maximum at $\p=\k/n$. Applying again the Laplace method we obtain: 
\begin{equation}
\mbox{P}_{\K}(k_1,\ldots,k_N)\approx\frac{\prod_{i=1}^N\left(\frac{k_i}{n}\right)^{-\frac{1}{2}}\delta(\sum_{i=1}^Nk_i-n)}{2^{N-1}\Omega},
\end{equation}
when $\frac{\k}{n}$ corresponds to a point in the domain $\mbox{\bf{D}}$; otherwise $\mbox{P}_{\K}(\k)=0$.  The individual information $\mbox{I}(\K;\p)$ can be conveniently evaluated by use of ``reduced'' distributions $\mbox{P}_{\K'/\P}(\k'/\p)$ and $\mbox{p}_{\K'}(\k)$, where we take into account explicitly the dependence between the components of the vector $\k$ implied by the $\delta$-function:
\begin{eqnarray} 
\mbox{P}_{\K'/\P}&=&\frac{\mbox{exp}[-\frac{1}{2}\sum_{i=1}^{N-1}\frac{(k_i-p_in)^2}{p_in}-\frac{\left(1-p_mn-\sum_{i=1}^{N-1}k_i\right)^2}{2p_mn}]}{(2\pi n)^{\frac{N-1}{2}}\prod_{i=1}^Np_i^{\frac{1}{2}}}\label{gaussian}\\
\mbox{P}_{\K'}(\k')&=&\frac{ \left(n-\sum_{i=1}^{N-1}k_i\right)^{-\frac{1}{2}}\prod_{i=1}^{N-1}{k_i}^{-\frac{1}{2}}} { 2^{(N-1)}\Omega n^{\frac{N}{2}-1}}
\end{eqnarray} 
Then
\[
\mbox{I}(\K;\p)=\mbox{I}(\K';\p)
\]
\[
=\int_{\frac{\k}{n}\in\mbox{\bf{D}}}\mbox{P}_{\K'/\P}(\k'/\p)\left[\ln\mbox{P}_{\K'/\P}(\k'/\p)-\ln\mbox{P}_{\K'}(\k')\right]dk_1\ldots dk_{N-1}
\]
\begin{equation}
=\mbox{I}_1+\mbox{I}_2.\label{two-terms}
\end{equation}
The first term in (\ref{two-terms}) is simply the differential entropy (with the opposite sign) of a multivariate ($N-1$)-dimensional Gaussian distribution (\ref{gaussian}) with the determinant of covariance matrix $d=n^{N-1}\prod_{i=1}^Np_i$. Hence
\begin{equation}
\mbox{I}_1=-\frac{1}{2}\ln\left[(2\pi e)^{N-1}d\right]=-\frac{1}{2}\ln\left[(2\pi en)^{N-1}\prod_{i=1}^Np_i\right]
\end{equation}
The second term in (\ref{two-terms}) can be evaluated by the Laplace method, since the integrand has a sharp maximum at $k_i=p_in$ $(i=1,\ldots,N-1)$. Hence
\begin{equation}
\mbox{I}_2=\ln\Omega+\ln(2n)^{N-1}+\frac{1}{2}\ln\prod_{i=1}^Np_i.
\end{equation}
Thus, $\mbox{I}(\K;\p)=\ln\Omega+\frac{N-1}{2}\ln\frac{2n}{\pi e}$ does not depend on $\p$.  This proves that the distribution (\ref{delta}) is the optimal one and the maximum information in $\K$ about $\P$ is asymptotically equal to 
\begin{equation}
\mbox{I}_{\mbox{\scriptsize sup}}(\K;\P)=\ln\Omega+\frac{N-1}{2}\ln\frac{2n}{\pi e}.
\end{equation}
The number of distinguishable states is given by the following expression:
\begin{equation}
\mbox{W}(N,n,\Omega)=\Omega\left(\frac{2n}{\pi e}\right)^{\frac{N-1}{2}}.\label{final-expression-general}
\end{equation}
Expression (\ref{final-expression-general}) turns into
(\ref{final-expression}) for $N=2$.  Indeed, it is easy to see that a uniform
2-dimensional distribution in Cartesian coordinates restricted to the
non-negative quadrant of a unit circumference results in a uniform
distribution over the polar angle $\alpha$.  Similarly, in the
$N$-dimensional case we obtain a uniform distribution over the area of the
domain $\mbox{\bf{D}}$, i.e. over the solid angle.  

The number of
distinguishable states reaches a maximum (for given $N$ and $n$) if
$\mbox{\bf{D}}$ is the entire non-negative orthant of the $N$-dimensional
unit sphere.  Since the area of the surface of the $N$-dimensional unit
sphere is $2\pi^{N/2}\left[\Gamma(N/2)\right]^{-1}$, the area of the
non-negative orthant (the solid angle) is
\begin{equation}
\Omega_{\mbox{\scriptsize max}}=\frac{\pi^{N/2}}{2^{N-1}\Gamma(N/2)},\label{omega-max}
\end{equation}
where $\Gamma$ is Euler's gamma-function.  Thus the maximum number of distinguishable states in $N$-dimensional space is
\begin{equation}
\mbox{W}_{\mbox{\scriptsize max}}(N,n)=\frac{\pi^{1/2}}{\Gamma(N/2)}\left(\frac{n}{2e}\right)^\frac{N-1}{2}.\label{w-max}
\end{equation}
Remember that (\ref{omega-max}) and (\ref{w-max}) are valid only when approximation (\ref{gnedenko-eqn}) is valid, i.e. when $n\gg N$.
\section{Conclusion}

The main result of the paper can be summarized as follows.  The number
of distinguishable quantum states in a $2$-dimensional Hilbert space
is proportional to the number of identical copies of each state to the power $\frac{N-1}{2}$ and to the area $\Omega$ of the domain of the unit sphere occupied by the state vectors. Surprisingly, it does not depend
on the shape and the position of this domain, provided that the main assumption $n/N\gg1$ is satisfied.  The domain does not have to be connected: the results hold for a set of separate domains with the same total area $\Omega$.  The optimal distribution is uniform over the domain, which suggests that the states should be chosen at equal angular distances from each other.  For the 2-dimensional case, the number of distinguishable states is proportional to the angular interval and to the square root of the number of identical copies of each state measured (cf. \cite{general-case}).

The result that the number of distinguishable states is proportional
to the geometric distance as measured by angle in Hilbert space is
quite nontrivial and noteworthy.  Indeed, it suggests that the metric
of Hilbert space may result not from a physical principle, but rather
as a consequence of an optimal statistical inference procedure.

\end{document}